# NDTAODV: Neighbor Defense Technique for Ad hoc On-Demand Distance Vector (AODV) to Mitigate Flood Attack in MANETs

Akshai Aggarwal[1], Savita Gandhi[2], Nirbhay Chaubey[2], Naren Tada[2], Srushti Trivedi[2]

[1]Gujarat Technological University, Ahmedabad – 380 015, Gujarat, India
[2]Department of Computer Science, Gujarat University, Ahmedabad – 380 009, Gujarat, India

*ABSTRACT*

*Mobile Ad Hoc Networks (MANETs) are collections of mobile nodes that can communicate with one another using multihop wireless links. MANETs are often deployed in the environments, where there is no fixed infrastructure and centralized management. The nodes of mobile ad hoc networks are susceptible to compromise. In such a scenario, designing an efficient, reliable and secure routing protocol has been a major challengesue over the last many years. The routing protocol Ad hoc On-demand Distance Vector (AODV) has no security measures in-built in it. It is vulnerable to many types of routing attacks. The flood attack is one of them. In this paper, we propose a simple and effective technique to secure Ad hoc On-demand Distance Vector (AODV) routing protocol against flood attacks. To deal with a flood attack, we have proposed Neighbor Defense Technique for Ad hoc On-demand Distance Vector (NDTAODV). This makes AODV more robust. The proposed technique has been designed to isolate the flood attacker with the use of timers, peak value and hello alarm technique.*

*We have simulated our work in Network Simulator NS-2.33 (NS-2) with different pause times by way of different number of malicious nodes. We have compared the performance of NDTAODV with the AODV in normal situation as well as in the presence of malicious attacks. We have considered Packet Delivery Fraction (PDF), Average Throughput (AT) and Normalized Routing Load (NRL) for comparing the performance of NDTAODV and AODV.*

*KEYWORDS*

*AODV; RREQ-Flood attack; NS-2.33; PDF; AT; NRL*

## 1. INTRODUCTION

Mobile ad-hoc Networks (MANETs) are collections of wireless mobile nodes, dynamically forming a temporary network without the use of any pre-defined network infrastructure or centralized administration. The functional challenge in the design of ad hoc networks is the development of dynamic routing protocols that can efficiently find routes between two communicating nodes. The routing protocol must be able to keep up with the high degree of node mobility that often changes the network topology drastically and unpredictably [1].

Many types of routing protocols in MANETs have been proposed and these protocols can be classified into the three categories: Table-driven (or Proactive), on-demand (or Reactive) and Zone based (or Hybrid). Many experiments have been done to find the best protocol out of the available protocols. Almost all experiments lead to the most used and reliable protocol namely Ad hoc On-demand Distance Vector (AODV), which is a reactive routing protocol. But the AODV protocol has no security measures in-built in it. Thus, it is vulnerable to many types of attacks.

In this paper, we demonstrate that the flood attack is possible in AODV routing protocol and propose techniques to provide security for preventing flood attacks. Our paper is divided into several sections as follows: Section 2 describes the fundamental working of AODV. In section 3, a description of the flooding attack is given. Section 4 discusses the attacker's approach in a flooding attack. Section 5 describes the previous work in this area. Section 6 provides complete understanding of our proposed algorithm - Neighbor Defense Technique for AODV (NDTAODV). Section 7 and 8 provide the simulation set up and result analysis respectively. We have concluded this paper in Section 9 followed by the references of our research work.

## 2. FUNDAMENTAL WORKING OF AODV

AODV routing protocol is an on-demand routing protocol where the route is established on demand. These kind of routing protocols compute the route to a specific destination only when it is needed. So a routing table, containing all the nodes as entries, does not have to be maintained in each node. When a source wants to send a packet to a destination, it invokes a route discovery mechanism to find the path to the destination. The route remains valid till the destination is reachable or until the route is no longer needed.

AODV consists of two processes- Route Discovery (use of RREQ and RREP) and Route Maintenance (use of RERR and HELLO). Figure 1 depicts the process of Route Discovery [3].

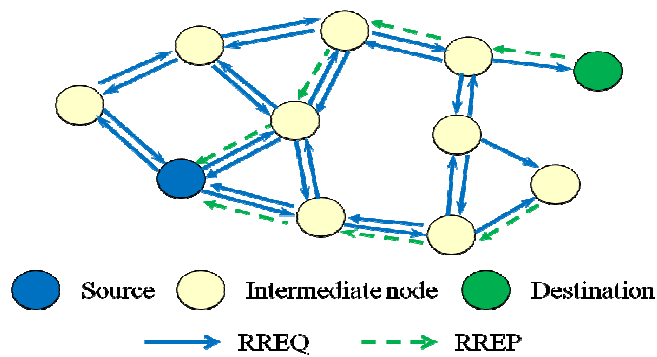

Figure 1. Fundamental Route Discovery Process in AODV

In AODV, route discovery is entirely on-demand. When a source node needs to send packets to a destination, for which it has no available route, it broadcasts a RREQ (Route Request) packet to its neighbors. Each node maintains a monotonically increasing sequence number to ensure loop free routing and supersedes stale route cache. The source node includes the known sequence number of the destination in the RREQ packet [3]. The intermediate node, receiving a RREQ packet, checks its route table entries. If it possesses a route toward the destination with a greater sequence number than that in the RREQ packet, it unicasts a RREP (Route Reply) packet back to its neighbor from which it received the RREQ packet. Otherwise, it sets up the reverse path and then rebroadcasts the RREQ packet. Duplicate RREQ packets received by one node are silently dropped. This way, the RREQ packet is flooded in a controlled manner in the network, and it will eventually arrive at the destination itself or at a node that can supply a fresh route to the destination, which will generate the RREP packet. As the RREP packet is propagated along the reverse path to the source, the intermediate nodes update their routing tables, with an additional constraint on the sequence number, and set up the forward path.

AODV also includes the route maintenance mechanism to handle the dynamics in the network topology. Link failures can be detected by either periodic beacons or link layer acknowledgments, such as those provided by 802.11 MAC protocol [3]. Once a link is broken, an unsolicited RRER packet with a fresh sequence number and infinite hop count is propagated to all active source nodes that are currently using this link. When the source node receives the notification of a broken link, it may restart the path discovery process if it still needs a route to the destination.

## 3. FLOODING ATTACK IN MANET

The malicious node(s) can attack in MANET using different ways, such as sending fake messages several times, fake routing information, and advertising fake links to disrupt routing operations [4]. In the following subsection, current routing attacks and its countermeasures against MANET protocols are discussed in brief.

### 3.1 Flooding Attack w.r.t to AODV

In flooding attacks, an attacker exhausts the network resources, such as bandwidth and to consume a node's resources e.g. computational and battery power or to disrupt the routing operation to cause severe degradation in network performance. In AODV protocol, a malicious node can send a large number of RREQs in a short period to a destination node that does not exist in the network. Because no one will reply to the RREQs, these RREQs will flood the whole network. As a result, all of the node battery power, as well as network bandwidth will be consumed and could lead to denial-of-service to other users.

## 4. ATTACKER APPROACH TO FLOOD THE NETWORK

The networks are particularly vulnerable to denial of service (DOS) attacks launched through compromised nodes or intruders. The new DOS attack, called Ad Hoc Flooding Attack can result in denial of service when used against on-demand routing protocols for mobile ad hoc networks, such as AODV/Dynamic Source Routing Protocol (DSR). The intruder broadcasts mass Route Request packets to exhaust the communication bandwidth and resources of nodes so that valid communication between nodes cannot be sustained. The injected packet is a fake packet. The attacker node puts its own define value in RREQ packet in order to make this attack more dangerous.

Flooding RREQ packets in the whole network will consume a lot of resources of the network. To reduce congestion in a network, the AODV protocol adopts the following method: It limits the number of messages,t originating from a node to RREQ_RATELIMIT RREQ messages per second. After broadcasting a RREQ, a node waits for a RREP. If a route is not received within round-trip milliseconds, the node may try again to discover a route by broadcasting another RREQ, up to a maximum of retry times at the maximum TTL value. In the Flooding Attack, the attack node violates the above rules to exhaust the network resource.

First, the attacker selects many IP addresses which are not in the network, if the attacker knows the scope of IP address in the networks. Because none of the nodes can answer RREP packets for these RREQ, the reverse route in the route table of node will be conserved longer. The attacker can select random IP addresses if it doesnot know the scope of IP addresses.

Secondly, the attacker successively originates mass RREQ messages for these void IP addresses. The attacker tries to send excessive RREQ without considering Request rate limit per second. The attacker will resend the RREQ packets without waiting for the RREP or round-trip time, if it uses out these IP addresses. The TTL of RREQ is set up to a maximum without using expanding ring search method. In the Flooding Attacks, the whole network will be full of RREQ packets which the attacker sends. The communication bandwidth is exhausted by the flooded RREQ packets and the resources of nodes are exhausted at the same time. For example, the storage of the route table is limited. If mass RREQ packets are coming to a node in a short time interval, the storage of route table in the node will be exhausted, so that the node would not be able to receive new RREQ packets. As a result, a legitimate nodes willnot be able to set up paths to send data. Figure 2 shows an example of RREQ Flooding Attack. Node 8 is an attacker and floods mass RREQ packets over the networks so that the other nodes cannot build paths with one another.

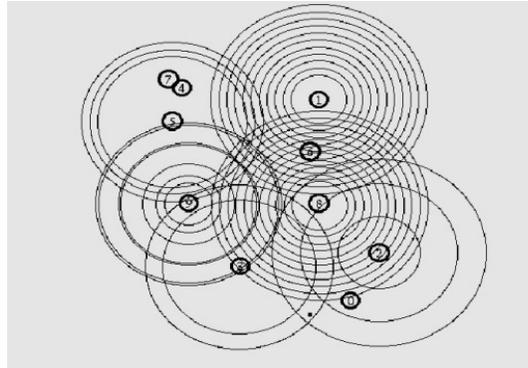

Figure 2. The Flooding Attack

## 5. PREVIOUS WORK

In [4] Rashid Hafeez Khokhar, Md Asri Ngadi and Satria Mandala reviewed routing attacks in MANETs. They have discussed various types of routing protocols and different kinds of possible attacks.

In [5] Ping Yi, Zhoulin Dai, Shiyong Zhang and Yiping Zhong proposed an approach to prevent flood attack and also compared Ad Hoc Flooding Attack and SYN Flooding Attack in AODV. Their results show that the packet delivery ratio becomes better compared to the one in original AODV.

In [6] Revathi Venkataraman, M. Pushpalatha, and T. Rama Rao proposed an algorithm to prevent flood attack. They define different trust levels for the neighbor, like a friend is most trusted, an acquaintances is less trusted and a stranger is least trusted or falls in the category of dangerous in terms of flood attack.

In [11] Akshai Aggarwal, Savita Gandhi etc, proposed AODVSEC wherein, they have modified the fundamental route discovery process of the basic AODV in such a way that the protocol can mitigate the effects of active forge attacks viz. Resource Consumption (RC) attack, Route Disturb (RD) attack and Route Invasion (RI) attack through fake RREP message. To provide the security provision, AODVSEC does not use cryptography or any central trusted authority, which may require a great deal of computational power. Performance of AODVSEC is no less than that of the SAODV but the same is achieved with lower processing requirement leading to saving of computational power. They have further proposed to focus on security aspects for the attacks that can be launched through forging the RREQ control message.

## 6. ALGORITHM FOR NEIGHBOR DEFENSE TECHNIQUE TO MITIGATE FLOOD ATTACK (NDT)

Flooding attack is very dangerous as far as MANET's performance is concerned. Such an attack adversely affects battery life and leads to congestion in the network. In this type of attack, requester (Attacker) floods the network with unnecessary RREQ packets. This can result into denial of service attack, as intermediate nodes have to do extra work to forward these fake packets. Thus the nodes may not be able to do other useful work and may not be able to involve itself in other operations. In our proposed algorithm, Broody list table and RREQ_count table are maintained by every node.

### 6.1 Broody list:

This table will keep the record for malicious node which floods the request. Flood Timer has taken for generating Dummy packet by the attackers. Every nodes maintain this table and keep the entries for those who are intruders.

Table 1. Broody List

| Malicious node 1 id |
|---|
| Malicious node 2 id |
| Malicious node 3 id |

### 6.2 RREQ_count table:

This table keeps track of the number of request come from each neighbor and    expiry value as timestamp in the particular interval. The Cache Timer will trigger the event for flushing the RREQ_count and check for the number of request comes from a neighbor if it is exceed the peak_value (that is the threshold value, which can be decided according to the network density), then it will place the requester in Broody list (dark-list) and will not allow requesters packet for broadcasting

Table 2.  RREQ_count

| RREQ_ID | RREQentry | TimeStamp |
|---|---|---|
| Requester1 Id | 5 | 0.34566 |
| Requester2 Id | 1 | 0.55346 |

**6.3 NDT Algorithm :**

```
if(CacheTimer trigger)
Then
    Flush RREQ_Count table entry
    if (check all entry for the RREQentry exceed the peak value in
    RREQ_Count table)
    Then
        Put the RREQester in broodyList!

if(RREQester is in broodyList)
Then
    Drop the packet

if ((RREQester is neighbor) && (there is no entry in RREQ_Count table))
Then
    Add the RREQentry for this RREQ in RREQ_Count table

if(RREQentry > PickValue)
Then
    Put the RREQester in broodyList!
```

Figure 3. The Neighbour Defence Technique (NDT) algorithm

Following are the Prerequisites and assumptions of NDTAODV:

As the detection can only done by the very neighbor of the attacker, enabling the HELLO packet of nodes will be required.

Peak_value, Cache interval and Flood interval should be synchronize with each other according to the nature of the MANET.

No other legitimate node sends more than 10 requests per second as they have to follows the AODV constraint. Attacker should be an outsider and so that it cannot block other nodes by entering other legitimate nodes in its broody list.

**6.4 HAT (Hello Alarm technique for global notification):**

In our work, we use the Hello alarm technique (HAT) for notification to other nodes in the network about existence of the malicious node. Our proposed algorithm utilizes the HELLO packet to transmit the entire Broody list to other legitimate nodes. Send hello and receive hello processes are used for retrieving and storing the malicious list from the packet.

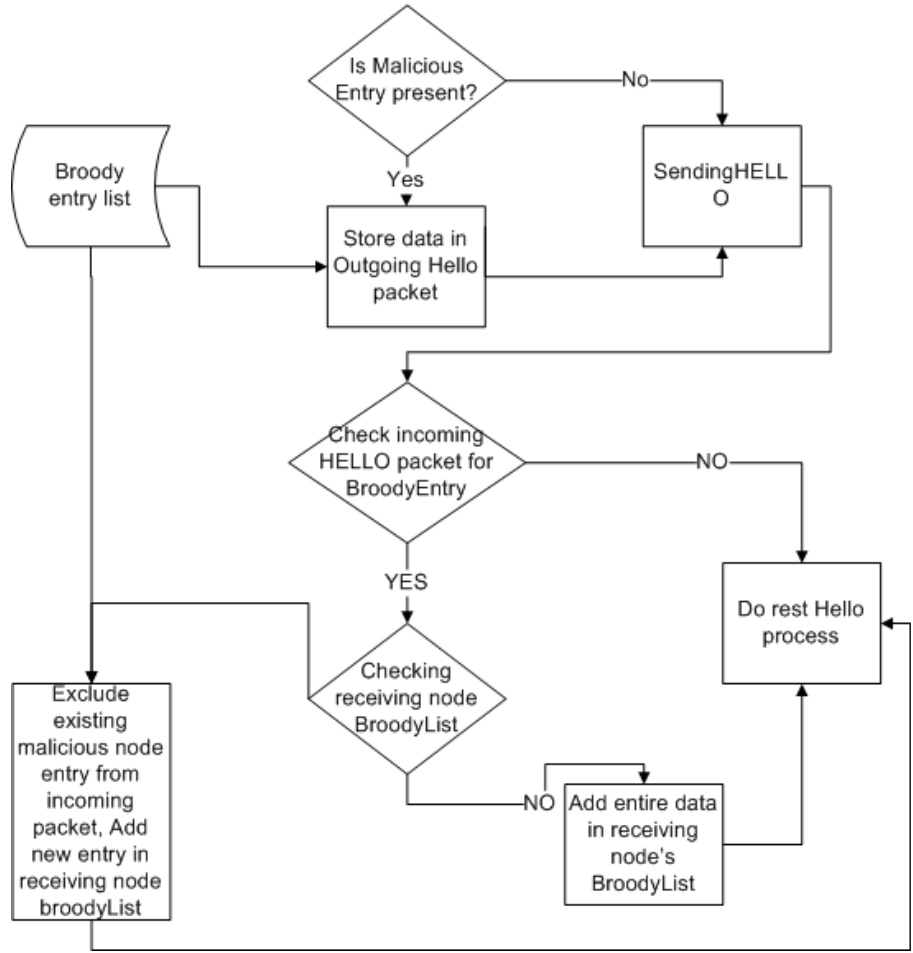

Figure 4. The Hello Alarm Technique (HAT) algorithm

## 7   SIMULATION SETUP

This section provides the simulation setup for comparing the performance of NDTAODV with the existing normal AODV using Network Simulator – 2.33 (NS-2) [7] [8]. The details of the timers and performance metrics are give in the following subsections. Table 3 shows details of simulation set up.

Table 2. Experimental Setup

| Parameter | Parameter Values |
|---|---|
| Simulator | NS-2.33 |
| Simulation time | 100 Second |
| Number of nodes | 25 |
| **Routing Protocol** | **AODV, NDTAODV** |
| **Traffic Model** | **CBR(UDP)** |
| **No. of sources/connections** | **5** |
| **Terrain area** | **1000 m X 1000 m** |
| **Mobility Model** | **Random Waypoint** |
| **MAC Protocol** | **IEEE 802.11** |
| **Antenna Type** | **Omni directional antenna** |
| **Propagation Model** | **Two Ray Ground** |
| **Packet Size** | **512 byte** |
| **Pause Time** | **0,5,10,15,20 ms** |
| **Number of Malicious Nodes** | **1,3** |
| **Flood Interval** | **0.009 Second** |
| **Cache Interval** | **1 Second** |
| **Peak Value** | **10 (Number of request)** |
| **Entry Expiry Time** | **CURRENT_TIME+1** |

### 7.1 Timers

Timers are default component in the AODV protocol. It will triggers specific event or function according to the timer value. The proposed algorithm two function constantly working SEND-fakerequest and ct_flush for floodTimer and cacheTimer respectively.

*FloodTimer:* In order to inject FAKE Request packet by a malicious node in the MANETs, Proposed scheme used flood timer which continuously sends the request packet as the value 0.009 second. Every 0.0009 second attacker broadcast the request packet in the network.

**CacheTimer:** In order to observe Request table entry for the expire time and request count entry for the requester to check whether it exceeds peak (Threshold) or not. The CacheTimer value set as 1.

### 7.2 Performance Metrics

Following three performance metrics are considered in our work:

(a) Packet Delivery Fraction (PDF): This is the ratio of the number of data packets successfully delivered to the destinations to those generated by sources.
(b) Average Throughput(AT): It is the rate of successfully transmitted data packets in a unit time in the network during the simulation.
(c) Normalized Routing Load (NRL): The number of routing packets transmitted per data packet delivered at the destination.

International Journal of Computer Networks & Communications (IJCNC) Vol.6, No.1, January 2014

## 8 RESULT ANALYSIS

This section discusses result of our experimental setup for three different cases. In this experimental work, we have considered network size of 25 nodes including 1 malicious node and five connections. Figures 5, 6 and 7 are the comparison graphs of PDF, Average Throughput and NRL vs Pause Time respectively. Different pause time means different mobility and pause time zero designated the highest mobility of nodes.

### 8.1 With flood attack for 1 malicious node

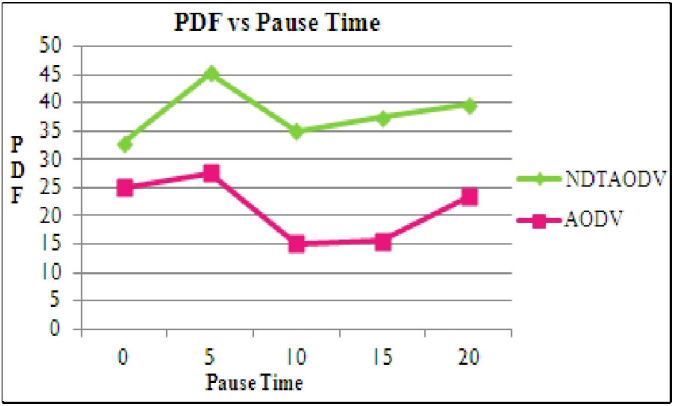

Figure 5. PDF vs PauseTime

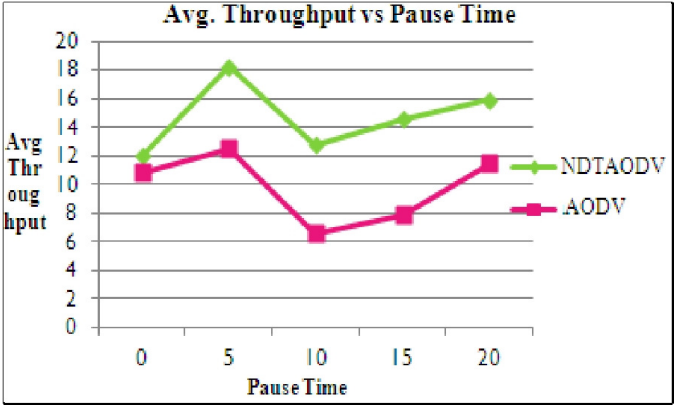

Figure 6. Average Throughput vs PauseTime

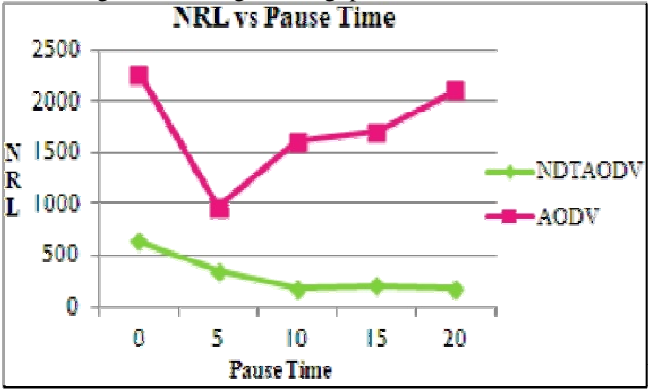

Figure 7. NRL vs PauseTime

27

Figure 5 describe that our work NDTAODV has   better packet delivery ratio than the normal AODV in the presence of flooding attack. Figure 6 show that the average throughput of NDTAODV is higher as compare to the original AODV and it continuously increased when the pause time increases in both the protocol. It is observed in the figure 7 that the normalized routing load (NRL) of NDTAODV is very much lower as compared to that of AODV

### 8.2 With  flood  attack  for  3  malicious node

In this experimental work, we have considered network size of 25 nodes including 3 malicious node and five connections. Figures 8, 9 and 10 are the comparison graphs of PDF, Average Throughput and NRL vs Pause Time respectively

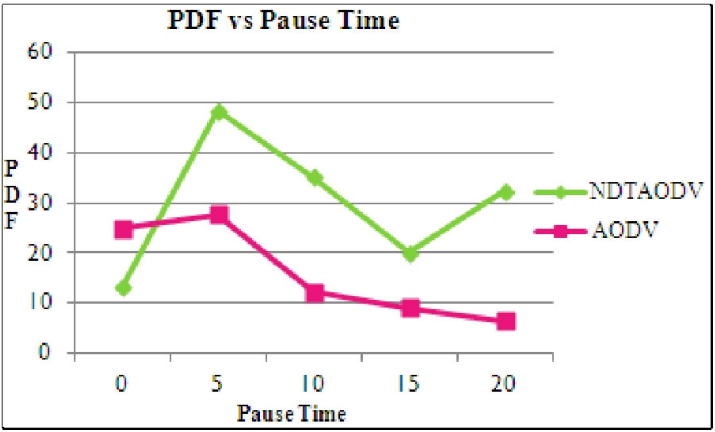

Figure 8. PDF vs PauseTime

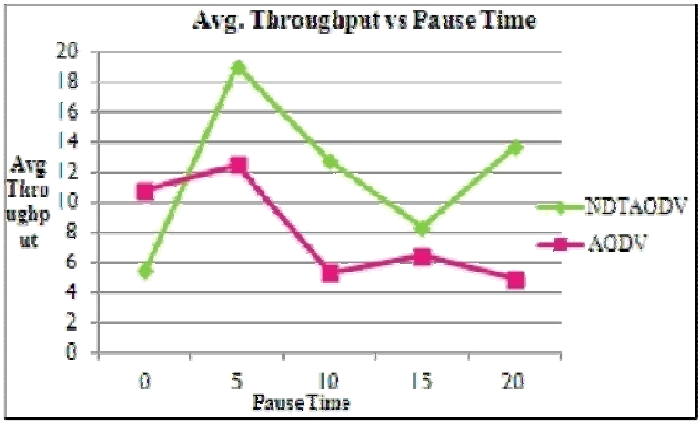

Figure 9. Average Throughput vs PauseTime

28

International Journal of Computer Networks & Communications (IJCNC) Vol.6, No.1, January 2014

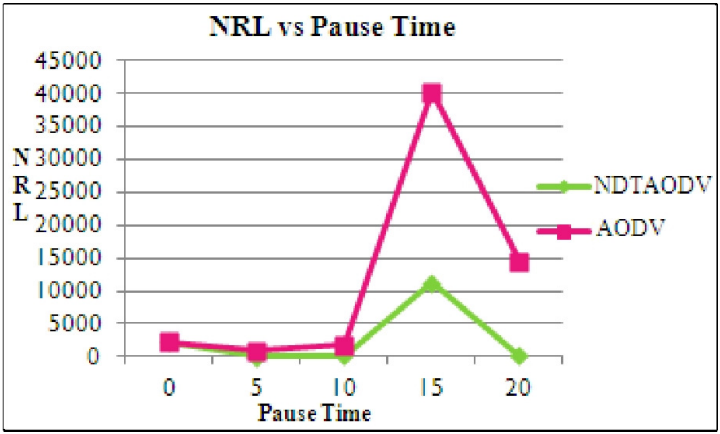

Figure 10. NRL vs PauseTime

Figure 8 shows that the PDF of NDTAODV is better even in the presence of three malicious nodes than that of normal AODV as long as mobility is not high. The performance of NDTAODV is reduced in high mobility (zero pause time) compared to the original AODV.

From the figure 9, we observe that the throughput of NDTAODV is less in high mobility compared to that of AODV but the throughput of our algorithm increased as and when pause time increased.

From Figure 10, we observe that NDTAODV and AODV both are having the same NRL in the high mobility. Further NRL of our proposed algorithm is continuously improved when the pause time is increased.

### 8.3 Without Flood attack under normal conditions

Following are the figures for comparison of Neighbor Defense Technique AODV (NDTAODV) and simple AODV in normal situation without any flood attack. We have taken a network of 25 nodes with five connection patterns. Figures 11, 12 and 13 represent the graphs of PDF, Average Throughput and NRL vs Pause Time respectively.

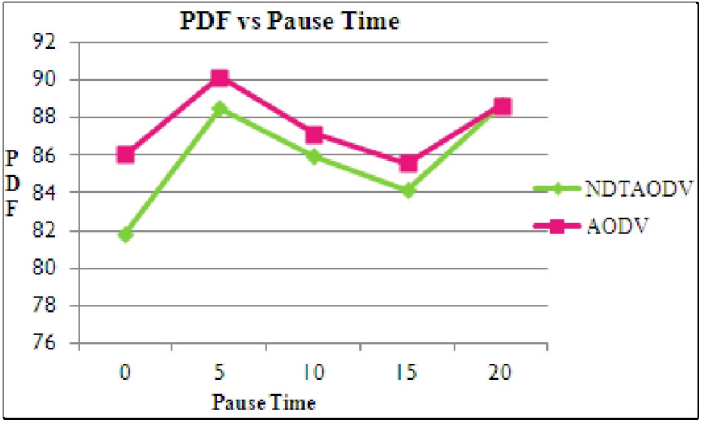

Figure 11. PDF vs PauseTime

29

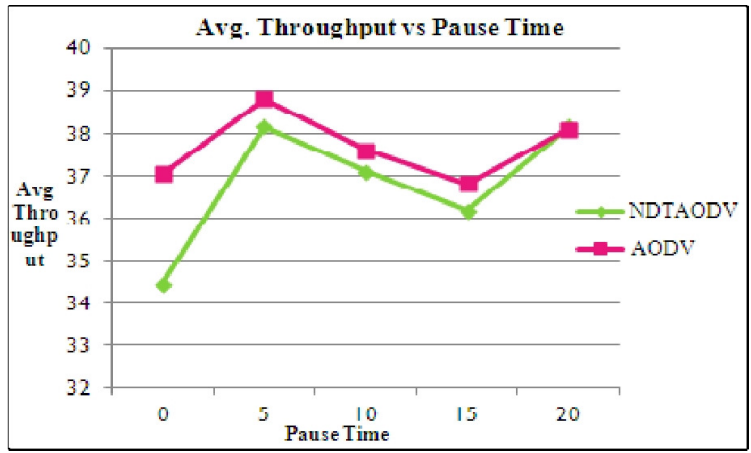

Figure 12. Average Throughput vs PauseTime

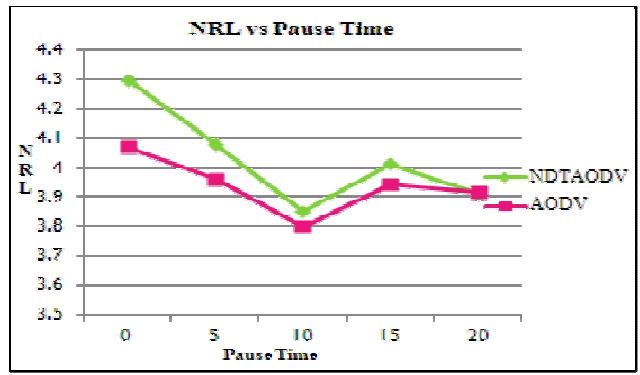

Figure 13. NRL vs PauseTime

From Figure 11, we observe that if no malicious node is present, then the PDF of our proposed algorithm NDTAODV is less compare to the PDF for the normal AODV in high speed mobility but after pause time of 15, our algorithm show a dynamic enhancement in the PDF continuously. As a result, we can conclude that NDTAODV works better for the military tank/application where mobility is less.

From Figure 12, we observe that in the absence of malicious activity, the throughput of our algorithm NDTAODV is less compared to that of the normal AODV for high mobility but after pause time 15, enhancement in throughput in NDTAODV is observed at a fast rate.

From Figure 13, we observe that the NRL of NDTAODV is higher compare to that of the normal AODV in high mobility but after pause time 15, NDTAODV has less routing load.

## 9   CONCLUSION

Security is an important aspect for deployment of Mobile Ad Hoc Networks. The route discovery process in the AODV is extremely susceptible to RREQ flooding attack, hence, it is imperative to provide an efficient security mechanism to mitigate the effect of such attacks. In this   paper, we propose an effective mechanism NDTAODV to provide security against RREQ flooding attacks in existing AODV routing protocol to establish a secure route for communication.

Using simulation results, the performance of NDTAODV is analyzed against RREQ flooding attack and justified. Our results show that the attacks have a great effect on the network performance and NDTAODV efficiently detects and isolate the malicious nodes from the active route to make the network available. Packet Delivery Fraction (PDF) of NDTAODV improve and Average Throughput (AT) which is the most important aspect of protocol is maintained as the pause time increase while AODV performance drops significantly under the presence of flood attack. So, performance of NDTAODV is better in the presence of attacks whereas normal AODV perform better in the absence of flooding attack when mobility increases. As a future work, we intend to incorporate the proposed scheme for presence of other type of attacks like greyhole attack, blackhole attack and wormhole attack which many affect the network performance.

International Journal of Computer Networks & Communications (IJCNC) Vol.6, No.1, January 2014

**Authors:**

**Akshai Aggarwal** (MIEEE'1966, SMIEEE'1992, LMIEEE 2011) is working as Vice Chancellor, Gujarat Technological University, Ahmedabad, India. Before joining as the Vice-Chancellor, he was working as the Director of School of Computer Science, University of Windsor, Canada. He worked as Professor and Head of Department of Computer Science at Gujarat University for about 10 years. Before that he was Professor and Head, Department of EE at M.S.University of Baroda. He was Chairman of IEEE India Council for two years. He initiated IEEE activities in Gujarat by 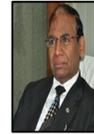 starting the first IEEE Student Branch at M.S.University of Baroda. Later he initiated the establishment of the Student Branch at Gujarat University. He was also the founder Chairman of IEEE Gujarat Section, the IEEE Computer Society  Chapter and the IEEE Joint Chapter of Industry Applications, Industrial Electronics and Power Electronics. The Section conducted two International Conferences and one national Seminar during his Chairmanship. He graduated with a B.Sc.(EE) from Punjab Engg College and studied at MS University of Baroda for his Master's and Doctoral work.

**Savita Gandhi** (MIEEE' 2003 SMIEEE' 2005) is Professor & Head at the          Department of Computer Science, Gujarat University and Joint Director, K.S. School of Business Management, Gujarat University. She is with Gujarat         University for about 24 years. Before that she has worked with M.S. University of Baroda, Department of Mathematics for about 10 years. She has been actively     associated with IEEE activities at Gujarat Section. She is M.Sc. (Mathematics), 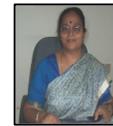 Ph.D (Mathematics) and A.A.S.I.(Associate Member of Actuarial Society of   India by the virtue of having completed the "A" group examinations comprising six subjects conducted by  Institute of Actuaries,      London). She was awarded Gold Medal for standing first class first securing 93% marks in M.Sc. and several prizes at M.Sc. as well as B.Sc. Examinations for obtaining highest marks.

**Nirbhay Chaubey** (SIEEE' 2002 MIEEE' 2004) pursuing his Ph.D (Computer Science) at Department of Computer Science, Gujarat University, Ahmedabad, India and working as an Assistant Professor of Computer Science at Institute of Science and Technology for Advanced Studies and Research, Vallabh          Vidyanagar, Gujarat, India. He has been involved in IEEE activities since 1994. His position held for IEEE Gujarat Section include Executive Secretary       (1998- 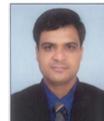 2005), Treasurer (2005-2006), Secretary and Treasurer (2007) and  Treasurer for year 2008 onwards. He graduated from Ranchi Unviersity, Ranchi, and Master in Computer Applications from Madurai Kamraj University, Madurai, India.

**Naren Tada** received M.Tech. (Networking & Communications) degree in year 2012 from Department of Computer Science,   Gujarat University, Ahmedabad. Currently, he is working as an Assistant Professor of Computer Engineering, V.V.P. Engineering College, Rajkot, Gujarat, India. 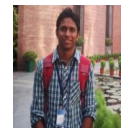

**Srushti Trivedi** received M.Tech. (Networking & Communications) degree in year 2012 from Department of Computer Science,   Gujarat University, Ahmedabad.  She opted the domain of Security Concerns in Mobile Ad hoc Networks.